\documentclass[numberedappendix,iop]{emulateapj}

\usepackage{dcolumn}
\usepackage{bm}
\usepackage{epsf}
\usepackage{verbatim}
\usepackage{amsmath}
\usepackage{amssymb}
\usepackage{color}
\usepackage{verbatim} 
\usepackage[normalem]{ulem} 
\usepackage{mathtools}
\usepackage{amsfonts}
\usepackage{relsize}

\usepackage{xcolor}
\definecolor{darkgreen}{rgb}{0.0,0.5,0.0}
\usepackage[colorlinks,citecolor=darkgreen]{hyperref} 

\usepackage[makeroom]{cancel}
\DeclareMathOperator*{\MyEqual}{=}
\DeclareMathOperator*{\MyEquiv}{\equiv}

\DeclareMathOperator*{\MySimeq}{\simeq}

\newcommand{\ie}{\emph{i.e.} }
\newcommand{\eg}{\emph{e.g.,} }

\newcommand{\be}{\begin{equation}}
\newcommand{\ee}{\end{equation}}
\newcommand{\bea}{\begin{equation*}}
\newcommand{\eea}{\end{equation*}}
\newcommand{\beq}{\begin{equation} }
\newcommand{\eeq}{\end{equation}}
\newcommand{\beqr}{\begin{eqnarray} \nonumber}
\newcommand{\eeqr}{\end{eqnarray}}
\newcommand{\beqrb}{\begin{eqnarray}}
\newcommand{\eeqrb}{\nonumber \end{eqnarray}}
\newcommand{\fin}{\mbox{ .}}
\newcommand{\coma}{\mbox{ ,}}

\newcommand{\const}{\mbox{const}}

\newcommand{\vect}[1]{\mathbf{#1}}

\newcommand{\pr}{\partial}

\newcommand{\lrgspc}{\,\,\,\,\,\,\,\,\,}
\newcommand{\smlspc}{\,\,\,\,}

\newcommand{\Gain}{g}
\newcommand{\MeanGain}{\llangle g\rrangle}
\newcommand{\MeanGainP}{\llangle 1+g\rrangle}

\newcommand{\myPesc}{P_{\tiny\text{esc}}}

\newcommand{\myPescIso}{P_{\tiny\text{esc,iso}}}
\newcommand{\iso}{{0}}

\newcommand{\sh}{{s}}
\newcommand{\st}{{\tiny\text{st}}}

\newcommand{\MyScat}{{\kappa}}
\newcommand{\MyScatB}{{w}}

\newcommand{\MyPsi}{{\phi}}
\newcommand{\MyPsiB}{{\psi}}
\newcommand{\Myd}{{d}}
\newcommand{\MyJ}{{j}}

\makeatletter
\DeclareFontFamily{OMX}{MnSymbolE}{}
\DeclareSymbolFont{MnLargeSymbols}{OMX}{MnSymbolE}{m}{n}
\SetSymbolFont{MnLargeSymbols}{bold}{OMX}{MnSymbolE}{b}{n}
\DeclareFontShape{OMX}{MnSymbolE}{m}{n}{
    <-6>  MnSymbolE5
   <6-7>  MnSymbolE6
   <7-8>  MnSymbolE7
   <8-9>  MnSymbolE8
   <9-10> MnSymbolE9
  <10-12> MnSymbolE10
  <12->   MnSymbolE12
}{}
\DeclareFontShape{OMX}{MnSymbolE}{b}{n}{
    <-6>  MnSymbolE-Bold5
   <6-7>  MnSymbolE-Bold6
   <7-8>  MnSymbolE-Bold7
   <8-9>  MnSymbolE-Bold8
   <9-10> MnSymbolE-Bold9
  <10-12> MnSymbolE-Bold10
  <12->   MnSymbolE-Bold12
}{}

\let\llangle\@undefined
\let\rrangle\@undefined
\DeclareMathDelimiter{\llangle}{\mathopen}%
                     {MnLargeSymbols}{'164}{MnLargeSymbols}{'164}
\DeclareMathDelimiter{\rrangle}{\mathclose}%
                     {MnLargeSymbols}{'171}{MnLargeSymbols}{'171}
\makeatother

\begin{document}

\bibliographystyle{apj}

\title{Diffusive shock-acceleration: breakdown of spatial diffusion and isotropy}

\shorttitle{breakdown of standard DSA}

\author{Uri Keshet}

\author{Ofir Arad}

\author{Yuri Lyubarski}

\shortauthors{Keshet et al.}

\affil{Physics Department, Ben-Gurion University of the Negev, POB 653, Be'er-Sheva 84105, Israel; ukeshet@bgu.ac.il}

\date{\today}

\begin{abstract}
We point out that particles accelerated in a non-relativistic shock of compression ratio $r$ attain the standard, $p=(r+2)/(r-1)$ spectral index only under certain conditions.
Previous derivations of the spectrum, based on the approximations of spatial diffusion or negligible anisotropy, are shown to be unjustified for a general scattering function.
We explain and demonstrate numerically that in contrast to previous claims, $p$ can substantially deviate from the standard result for anisotropic scattering.
We prove analytically that the standard approach is nevertheless valid in the limit of an isotropic medium.
Additional spectral modifications, for example by motions of scattering modes at intermediate optical depths from the shock, are discussed.
\end{abstract}


\maketitle

\section{Introduction}
\label{sec:Intro}

Diffusive shock acceleration (DSA) is a first-order Fermi process, believed to be responsible for the production of non-thermal, high-energy distributions of charged particles in collisionless shocks found in diverse astronomical systems.
For reviews, see \citet{Blandford_Eichler_87, MalkovDrury01, TreumannEtAl09Review}.
DSA is thought to operate in both non-relativistic and relativistic shocks, the latter being complicated by substantial anisotropy and sensitivity to microphysical processes \citep[\eg][]{BykovEtAl12Review, SironiEtAl15_review}.
We focus here on DSA in the non-relativistic shock limit, in which the shock-frame fluid velocity $v$ normalized by the speed $c$ of light, $\beta\equiv v/c\ll1$, is a small parameter.

Collisionless shocks in general, and their particle acceleration in particular, are mediated by electromagnetic modes, and are still not generally understood from first principles. No present analysis self-consistently calculates the long-term generation of these modes and their cross-interactions with the multi-phase plasma.
One way to make progress in the study of this non-linear, many body, and multi-scale problem is to evolve the particle distribution function (PDF) $f$ by adopting some ansatz for the scattering mechanism and neglecting wave generation and shock modification by the accelerated particles, in the so-called test-particle approximation.

This approach proved successful in accounting for observations of nonthermal shock signatures. For non-relativistic shocks, DSA is thought to yield a power-law energy spectrum, $n(E)\propto E^2 f\propto E^{-p}$, with a spectral index
\begin{equation} \label{eq:DSAp}
p\simeq p_0\equiv \frac{r+2}{r-1}
\end{equation}
that is a function of the shock compression ratio $r$ \citep{Krymskii_1977, AxfordEtAl77, Bell_1978, BlandfordOstriker78}. For a strong shock in an ideal gas of adiabatic index $\Gamma=5/3$, the compression ratio $r\to4$ implies $p\to2$, in agreement with observations.
Equation (\ref{eq:DSAp}) is often used to deduce $r$ --- and therefore, also the shock Mach number --- from the particle spectrum inferred from observations.

Analytic estimates of $p$ leading to Eq.~(\ref{eq:DSAp}) are typically based on applying the spatial diffusion approximation on both sides of the shock, or on approximating the PDF in the downstream frame as isotropic.
Indeed, spatial diffusion is a good approximation far from the shock, where gradients become small.
Similarly, a low, $O(\beta)$ level of anisotropy is expected due to the small probability of particles to escape downstream of a non-relativistic shock.
Moreover, Eq.~(\ref{eq:DSAp}) was confirmed by a wide range of numerical \citep[\eg][]{EllisonEtAl90, BednarzOstrowski98, Kirk_2000} and semi-analytic \citep{Keshet06} methods.

The standard lore is that the spectrum (\ref{eq:DSAp}) is guaranteed in the test-particle approximation for an arbitrary small-angle \citep[\eg][]{MalkovDrury01} or even large-angle \citep[\eg][]{BlasiVietri05} scattering function, independent of any first order anisotropy pattern that may emerge \citep[\eg][]{vietri2008Book}.
However, the result was not rigorously proven, to our knowledge, without invoking spatial diffusion or downstream isotropy, explicitly or implicitly.
It is therefore necessary to critically examine the spectrum in the presence of subtle effects that deviate from spatial diffusion, and taking into account small anisotropies.
Such an examination is needed in order to determine the circumstances under which Eq.~(\ref{eq:DSAp}) breaks down, and to rigorously establish it where it holds.

The spatial diffusion approximation is analyzed in \S\ref{sec:SpatialDiffusion}.
We argue (in \S\ref{sec:NoSpatialDiff}) that deviations from spatial diffusion in the vicinity of the shock front, where particle streaming and rapid changes in the scattering-mode properties may become substantial, must be carefully dealt with.
Computing $p$ by applying the spatial diffusion approximation near the shock front is self-consistent only under a certain condition (derived in \S\ref{sec:SpatialDiffAs}), satisfied under special circumstances such as an isotropic medium.

The general problem of arbitrarily large-angle scattering is studied in \S\ref{sec:LargeAngle}.
First, we analytically solve (in \S\ref{sec:LargeAngleIso}) the simple case of DSA with isotropic scattering, which was not previously addressed to our knowledge.
The general case is then analyzed (in \S\ref{sec:LargeAngleGen}); a modified diffusion equation is derived, and deviations from the spectrum (\ref{eq:DSAp}) are quantified and shown to become substantial for sufficiently anisotropic scattering.

Focusing on the limit of small-angle scattering (in \S\ref{sec:SmallAngle}), we show (in \S\ref{sec:Anisotropy}) that the shock-front PDF has an angular derivative that is first order in $\beta$ even in the downstream frame, and (in \S\ref{sec:AnisotropyImportant}) that a general $O(\beta^1)$ anisotropy is sufficiently strong to substantially distort the spectrum (\ref{eq:DSAp}).
The conditions necessary for the spectrum to deviate from Eq.~(\ref{eq:DSAp}) are quantified (in \S\ref{sec:pDev}), and shown (in \S\ref{sec:DiffExamples}) to be realized for sufficiently anisotropic angular diffusion.
Our results are summarized and discussed in \S\ref{sec:Discussion}.

\section{The assumption of spatial diffusion}
\label{sec:SpatialDiffusion}

Most derivations of Eq.~(\ref{eq:DSAp}) assume, directly or implicitly, that the evolution of relativistic particles around the shock can be approximated as a combination of advection and spatial diffusion. Here we discuss this approach, its underlying assumptions, and its breakdown.
An alternative approach, deriving the spectrum by neglecting the small anisotropy of particles downstream, is deferred to the study of small-angle scattering in \S\ref{sec:SmallAngle}.

\subsection{Breakdown of spatial diffusion}
\label{sec:NoSpatialDiff}

In the spatial diffusion approximation, one typically works in the shock frame, where a steady state is assumed to form.
Positing that the particle distribution is nearly isotropic, it is common to invoke the advection-diffusion equation
\begin{equation} \label{eq:DiffEq0}
0=\frac{\partial \tilde{N}}{\partial t} + \frac{\partial \tilde{\MyJ}}{\partial z} = c\beta \frac{\partial \tilde{N}}{\partial z} - \frac{\partial}{\partial z}\left[\mathcal{D}(z)\frac{\partial \tilde{N}}{\partial z}\right]
\end{equation}
separately on each side of the shock, and match the solutions at the shock front, $N_u(z=0)=N_d(z=0)$.
Here, $z$ is the displacement from the shock, upstream ($z<0$; subscript $u$) or downstream ($z>0$; subscript $d$), $N(z)$ is the particle number density, $\MyJ(z)=-\mathcal{D}\partial_z N$ is the diffusive particle flux, $\mathcal{D}(z)$ is the spatial diffusion function, and shock-frame parameters are designated by a tilde (omitted when unnecessary).

Solving Eq.~(\ref{eq:DiffEq0}) under the constraint that no particles escape upstream, $N(z\to-\infty)=0$, then implies that the particle density is uniform downstream, $N_d(z)=\const$.
Under such circumstances, Eq.~(\ref{eq:DSAp}) follows directly, as shown in several methods \citep{Krymskii_1977, AxfordEtAl77, Bell_1978, BlandfordOstriker78} and reproduced below.
However, this approach is valid only if deviations from the spatial diffusion approximation, which may become substantial near the shock front, remain both weak and confined to the close vicinity of the shock, at distances much smaller than the $\mathcal{D}/v$ diffusive scale.

Spatial diffusion is a good approximation only at large optical depths from an anisotropic source.
Defining $\tau=\tau(z)$ as the optical depth from the shock front (with $\tau<0$ upstream), one can safely invoke spatial diffusion at $|\tau|\gg1$.
In general, however, the approximation breaks down in the close vicinity of the shock front, $|\tau|\lesssim1$, where particles streaming across the shock can induce anisotropy patterns and spatial gradients that are inconsistent with spatial diffusion.
The problem is exacerbated in circumstances under which one cannot even sharply define $\tau(z)$, as we discuss in \S\ref{sec:LargeAngleGen}.

In \S\ref{sec:SpatialDiffAs}, we derive a consistency requirement on the angular distribution of the accelerated particles near the shock front, which must be satisfied if the spatial diffusion approximation is to be used to derive the spectrum.
Furthermore, we show that this requirement is violated for a general particle scattering function, necessitating a more careful approach.
Before proceeding with the formal analysis, we make some general comments to argue that deviations from spatial diffusion are to be expected, and may well modify the particle spectrum.

First, notice that some particle streaming must be present near the shock front.
Invoking spatial diffusion both upstream and downstream yields a discontinuous diffusive flux $\MyJ$ across the shock front.
In particular, particles crossing from the upstream induce a non-vanishing flux in the downstream frame, whereas $\MyJ_{d}\propto \partial_z N_d=0$.
A discontinuous diffusive flux does not directly pose any contradiction; in fact, the compensating flux term due to shock compression has been used \citep{Krymskii_1977} to derive the spectrum.
However, a discontinuous $\MyJ$ does indicate that the angular distribution may not be consistent with spatial diffusion on both sides of the shock, as we quantify in \S\ref{sec:SpatialDiffAs}.

We also show (therein) that, in the spatial diffusion approximation, the particle anisotropy downstream must satisfy a constraint, which is automatically fulfilled only if this anisotropy is $O(\beta^2)$.
However, particle streaming across the shock can induce an uncontrolled first-order downstream anisotropy.
Such an anisotropy is sufficiently strong to invalidate the standard spectrum (\ref{eq:DSAp}), as we prove in \S\ref{sec:AnisotropyImportant}.
To see how such a situation could arise, note that a first order anisotropy would imply a downstream streaming velocity of order $\beta$ near the shock.
If such streaming persists over distances $\sim\mathcal{D}/v$, it can induce order-unity variations in $N_d(z)$, which would necessarily invalidate Eq.~(\ref{eq:DSAp}).

In order to further illustrate the importance of streaming-like effects, consider a simplified form of particle streaming added to the diffusion equation,
\begin{equation} \label{eq:DiffEq}
c\beta \frac{\partial \tilde{N}}{\partial z} - \frac{\partial}{\partial z}\left[\mathcal{D}\frac{\partial \tilde{N}}{\partial z}\right] + \frac{\partial}{\partial z}\left[v_{\st}(z)\tilde{N}\right] = 0 \coma
\end{equation}
where the form of the last term is justified in \S\ref{sec:LargeAngleGen}.
The spectrum inferred from this equation would in general depend on the streaming velocity $v_{\st}(z)$,
and could significantly deviate from the standard Eq.~(\ref{eq:DSAp}).

As a concrete, analytically tractable example, consider downstream streaming of the form $v_{\st}(z)=v_0/[1+(z/z_0)]$, where $v_0$ and $z_0$ are constants, and define a dimensionless streaming length parameter $\alpha\equiv (z_0\beta c/\mathcal{D})$.
Equation~(\ref{eq:DiffEq}) yields $\tilde{N}(z\to\infty)/\tilde{N}(0)=\alpha_d^{-1}e^{-\alpha_d}E_{\alpha_0}(\alpha_d)$, where $E_a(b)$ is the exponential integral function.
Unless $|\alpha_0|\ll 1$, this ratio substantially differs from unity, so the resulting spectral index \citep[computed for example in the method of][as outlined in \S\ref{sec:AnisotropyImportant}]{Bell_1978}, $p=1+3\alpha_d^{-1}e^{-\alpha_d}E_{\alpha_0}(\alpha_d)/(r-1)$, similarly deviates from $p_0$.
In this picture, the standard spectrum would require the streaming term to be both small and confined to the close vicinity of the shock, $z_0=\alpha_0\mathcal{D}/v_0\ll\mathcal{D}/(\beta c)$.

Additional, independent caveats in the standard spectral derivations stem from the rather strong underlying assumptions, such as neglecting variations in $v=v_z$ and in $\mathcal{D}$ that may take place over length scales shorter than the spatial diffusion length.
For instance, consider a precursor to the shock upstream, or an offshoot downstream, in which the velocity has not yet reached its saturation value implied by the Rankine-Hugoniot adiabat, or an effective cutoff on $\mathcal{D}$ at some finite distance from the shock.
Such effects would modify the spectrum, for example by effectively altering the value of $r$ used in Eq.~(\ref{eq:DSAp}).
An interesting example is the backreaction of the accelerated particles on $\mathcal{D}$, which could severely modify the spectrum, as suggested by non-linear DSA studies \citep{drury_83, Blandford_Eichler_87, JonesEllison91, MalkovDrury01}.

In this paper, we adhere to the standard assumptions of a uniform fluid velocity on each side of the shock, and a simple scattering function that is not inherently high-dimensional (for example, has a separable angular dependence).
After establishing in \S\ref{sec:SpatialDiffAs} that even in this case, the spatial diffusion approximation is in general insufficient for a self-consistent derivations of the spectrum, we invoke the transport equation, and consistently solve for the full PDF in space and in momentum.
In general, this can only be done numerically, as we demonstrate in \S\ref{sec:LargeAngleGen} and \S\ref{sec:DiffExamples}.
Some analytical results can however be obtained; in particular, we show (in \S\ref{sec:LargeAngleGen}) that the diffusion equation must be replaced by a modified, diffusion-like equation, supplemented by both a streaming term and an anisotropy-dependent diffusive term.

\subsection{Spatial diffusion: consistency requirement}
\label{sec:SpatialDiffAs}

Consider an infinite, planar shock front, located at shock-frame coordinate $z=0$, with flow in the positive z direction both upstream ($z<0$) and downstream ($z>0$). We analyze the PDF $f(\vect{r},\vect{q},t)$ of energetic particles, crossing the shock back and forth due to repeated elastic scattering events that modify the direction $\vect{\hat{q}}$ of the fluid-frame momentum $\vect{q}$ of a particle.
Particle scattering can be parametrized in terms of an effective scattering function $\MyScat(\vect{\hat{q}},\vect{\hat{q}}')$, describing the fluid-frame rate at which particles moving in direction $\vect{\hat{q}}'$ are scattered to direction $\vect{\hat{q}}$.

Assuming that the PDF is stationary in the shock frame, $f$ is then governed by the transport equation, which can be conveniently written in the form \citep[\eg][]{Vietri03}
\begin{align}\label{eq:transport_kappa}
& (\mu_i+\beta_i)c\gamma_i \partial_z f(z,q_i,\vect{\hat{q}}_i) = \\
& \int \left[\MyScat_i(\vect{\hat{q}}_i,\vect{\hat{q}}')f(z,q_i,\vect{\hat{q}}')-
\MyScat_i(\vect{\hat{q}}',\vect{\hat{q}}_i)f(z,q_i,\vect{\hat{q}}_i)\right]\gamma_i\,d\Omega' \coma \nonumber
\end{align}
in the mixed phase space of fluid-frame $\vect{q}$ and shock-frame $z$.
Here, $\gamma\equiv(1-\beta^2)^{-1/2}$ is the fluid Lorentz factor, $\mu\equiv \vect{\hat{q}}\cdot \vect{\hat{z}}$ is the cosine of the polar angle with respect to the shock normal,
$d\Omega$ is the solid angle differential, and integrals are taken over the full domain unless otherwise specified.
Such an equation holds independently upstream and downstream of the shock, with the $\{\mbox{upstream, downstream}\}$ indices $i\in\{u,d\}$ written henceforth only when necessary.
Focusing on non-relativistic shocks, we henceforth approximate $\gamma\to1$.

Note that it is tacitly assumed here that by averaging $f(\vect{r},\vect{q},t)$ over constant $z$ planes, one arrives at a well-defined, lower-dimensional PDF, that in a steady state may be written as $f(z,q,\vect{\hat{q}})$.
It is also assumed that the functions $\MyScat_i$ are spatially uniform on each side of the shock, although this assumption is relaxed in \S\ref{sec:LargeAngle}.
These, and other underlying, standard assumptions, are discussed in \S\ref{sec:Discussion}.

It follows from the unitarity of the scattering matrix (e.q., \citealt{PhysKinetics81} section 2), that
\begin{equation} \label{eq:Unitarity}
\int \MyScat(\vect{\hat{q}},\vect{\hat{q}}')d\Omega'=\int \MyScat(\vect{\hat{q}}',\vect{\hat{q}})d\Omega' \fin
\end{equation}
Therefore, one can generally rewrite Eq.~(\ref{eq:transport_kappa}) as
\begin{equation}\label{eq:transport0}
\!\!\!(\mu+\beta)c\partial_z f =
\int \MyScat(\vect{\hat{q}},\vect{\hat{q}}')\left[f(z,q,\vect{\hat{q}}')-
f(z,q,\vect{\hat{q}})\right]\,d\Omega' .
\end{equation}
We avoid the stronger condition of detailed balance,
$\MyScat(\vect{\hat{q}},\vect{\hat{q}}')=\MyScat(\vect{\hat{q}}',\vect{\hat{q}})$,
which is sometimes invoked here \citep[\eg][]{BlasiVietri05}, because it would require time reversibility and parity symmetry of the scattering process, which could be violated in a magnetized medium.

As ultra-relativistic particles of energy $E$ much higher than any characteristic scale in the problem are expected to form a power-law energy spectrum, we separate variables by defining
\begin{equation} \label{eq:PhiDef}
f(z,q,\vect{\hat{q}})\equiv \MyPsi(z,\vect{\hat{q}}) q^{-(2+p)} \fin
\end{equation}
Then the transport equation is reduced to
\begin{equation}\label{eq:transport_psi}
\!\!\!(\mu+\beta)c \partial_z \MyPsi =
\int \MyScat(\vect{\hat{q}},\vect{\hat{q}}')\left[\MyPsi(z,\vect{\hat{q}}')-
\MyPsi(z,\vect{\hat{q}})\right]\,d\Omega' \fin
\end{equation}
To determine the spectral index $p$, we next incorporate the boundary conditions.

Continuity across the shock front implies that
\beq \MyPsi_u(z=0,\vect{\hat{q}}_u)q_u^{-(2+p)} = \MyPsi_d(z=0,\vect{\hat{q}}_d)q_d^{-(2+p)} \coma \label{eq:continuity} \eeq
where upstream and downstream quantities are related by a Lorentz boost of velocity $\beta_r=(\beta_u-\beta_d)/(1-\beta_u \beta_d)$, so
$q_d=(1+\beta_r \mu_u)\gamma_r q_u$ and $\mu_d=(\mu_u+\beta_r)/(1+\beta_r \mu_u)$.
In the non-relativistic limit, $\beta\ll 1$, Eq.~(\ref{eq:continuity}) reduces to
\begin{align}
\label{eq:Jump}
\MyPsi_{d}(z=0,\vect{\hat{q}}_d)
& = [(1+\beta_r \mu_u)\gamma_r]^{2+p} \MyPsi_{u}(z=0,\vect{\hat{q}}_u) \\
& \simeq [1+(2+p)(\beta_u-\beta_d)\mu_u] \MyPsi_{u}(z=0,\vect{\hat{q}}_u) \fin \nonumber
\end{align}
The escape of particles downstream and their absence far upstream imply that
\beq \lim_{z \to -\infty} \MyPsi_u = 0 \lrgspc \mbox{and} \lrgspc  \lim_{z \to +\infty} \MyPsi_d = \phi_\infty \coma \label{eq:boundary} \eeq
where $\MyPsi_\infty$ is an arbitrary constant, which we choose to be of order unity.

Averaging the transport equation (\ref{eq:transport_psi}) over angles $\vect{\hat{q}}$, and introducing the notation $\langle\dots\rangle\equiv (4\pi)^{-1}\int\dots d\Omega$, one arrives at the first integral of the transport equation,
\beq
\frac d{dz}\left(\beta\langle\phi\rangle+\langle\mu\phi\rangle\right)=0 \fin
\eeq
Applying the boundary conditions (\ref{eq:boundary}) now yields
\begin{equation} \label{eq:meanTrans}
\beta \langle\phi\rangle + \langle \mu \phi \rangle =
\begin{cases}
\beta_d \phi_\infty & \mbox{for $z>0$ ;} \\
0 & \mbox{for $z<0$ .}
\end{cases}
\end{equation}
Upstream, the anisotropy measure $\langle \mu \phi \rangle/\langle\phi\rangle$ is therefore small, of order $\beta$.
Assuming that $\phi_\infty/\langle\phi\rangle$ does not greatly exceed unity, $\langle \mu \phi \rangle/\langle\phi\rangle$ is small also downstream, of order $O(\beta)$.
It is thus useful to represent the distribution function as
\beq
\phi(z,\vect{\hat{q}})=\phi_\iso(z)+\phi_1(z,\vect{\hat{q}}) \coma
\label{eq:diffus}\eeq
where $\phi_\iso>0$, with the normalization
\beq \label{eq:PhiOneNorm}
\langle\phi_1\rangle=0 \fin
\eeq

Substituting Eq.~(\ref{eq:diffus}) into the transport equation (\ref{eq:transport_psi}) now yields
\beq \label{eq:transportKappa2}
\!\!(\mu+\beta)c\partial_{z}\phi=\int \MyScat(\vect{\hat{q}},\vect{\hat{q}}')\left[\phi_1(z,\vect{\hat{q}}')-
\phi_1(z,\vect{\hat{q}})\right]d\Omega'\coma
\eeq
to be solved for $\phi_\iso$ and $\phi_1$ on each side of the shock under the boundary conditions (\ref{eq:boundary}).
These solutions must then be matched using the continuity equation (\ref{eq:Jump}).
One may usually assume that $|\phi_1|\ll \phi_0$, although this assumption can locally break down for extreme scattering functions, as we show in \S\ref{sec:DiffExamples}.
Under this assumption, Eq.~(\ref{eq:Jump}) reduces to lowest order to
\beq
\phi_{u\iso}+\phi_{u1}+(2+p)(\beta_u-\beta_d)\mu\phi_{u\iso}=\phi_{d\iso}+\phi_{d1} \coma
\label{eq:JumpDiff}\eeq
evaluated at $z=0$.
Averaging over angles yields
\beq\phi_\iso=\phi_{u\iso}=\phi_{d\iso} \coma
\label{eq:JumpDiff0}
\eeq
and so
\beq
\phi_{u1}+(2+p)(\beta_u-\beta_d)\mu\phi_{u\iso}=\phi_{d1} \fin
\label{eq:JumpDiff2}\eeq

The small anisotropy parameter, $\langle \mu \phi \rangle/\langle\phi\rangle\ll 1$, found upstream and usually also downstream, motivates the aforementioned spatial-diffusion approximation. A sufficient condition for this approximation is a highly homogeneous or small level of anisotropy, in the sense that $|\partial_z \phi_1|\ll |\partial_z\phi_\iso|$. If, in addition, the gradient of the density is small, then the particle distribution is guaranteed to be nearly isotropic, $|\phi_1|\ll \phi_0$, as we show below.
Later, in \S\ref{sec:SmallAngle}, we show that both of these assumptions can be broken in an anisotropic medium.
Nevertheless, for the remainder of \S\ref{sec:SpatialDiffAs}, we adopt these assumptions, and examine their consequences.

Under the assumption $|\partial_z \phi_1|\ll |\partial_z\phi_\iso|$, retaining the lowest order (in $\beta$) terms in Eq.~(\ref{eq:transportKappa2}) yields
\beq \label{eq:transportKappa3}
\mu c\partial_{z}\phi_\iso=\int \MyScat(\vect{\hat{q}},\vect{\hat{q}}')\left[\phi_1(z,\vect{\hat{q}}')-
\phi_1(z,\vect{\hat{q}})\right]d\Omega'.
\eeq
One can separate the angular and spatial variables by defining a single-variable function $\MyPsiB(\vect{\hat{q}})$ through
\beq \label{eq:PsiDef}
\phi_1(z,\vect{\hat{q}})=\MyPsiB(\vect{\hat{q}}) c \partial_{z}\phi_\iso \coma
\eeq
thus reducing the problem to solving the integral equation
\beq
\int \MyScat(\vect{\hat{q}},\vect{\hat{q}}')\left[\MyPsiB(\vect{\hat{q}}')-
\MyPsiB(\vect{\hat{q}})\right]d\Omega'=\mu \label{eq:psi}
\eeq
for $\MyPsiB$.
Once this equation is solved, the particle flux may be written in the form
\beq
\langle\mu\phi\rangle=\langle\mu\phi_1\rangle=-\frac{\mathcal{D}}{c}\partial_{z}\phi_\iso \coma
\label{eq:Fick}\eeq
where the spatial diffusion coefficient is defined as
\beq \label{eq:DDef}
\mathcal{D}\equiv-c^2\langle\mu\MyPsiB\rangle \coma
\eeq
and in the present framework is a constant on each side of the shock.
A small density gradient, in the sense that $|\partial_z \phi_0|\ll c\phi_0/\mathcal{D}$, would now imply a nearly isotropic distribution, $|\phi_1|\ll \phi_0$.
In order to derive the spectrum --- and even the function $\phi_{d}(z)$ --- it is not necessary to determine $\psi$ or $\mathcal{D}$; rather, it suffices to assume that a consistent solution to Eq.~(\ref{eq:psi}) exists.

Indeed, substituting the spatial-diffusion law (\ref{eq:Fick}) into Eq.~(\ref{eq:meanTrans}) yields a closed equation for the particle density,
\begin{equation} \label{eq:meanTrans2}
\beta \phi_\iso - \frac{\mathcal{D}}{c}\partial_z\phi_\iso =
\begin{cases}
\beta_d \phi_\infty & \mbox{for $z>0$ ;} \\
0 & \mbox{for $z<0$ .}
\end{cases}
\end{equation}
Combining the solution to Eq.~(\ref{eq:meanTrans2}) with Eq.~(\ref{eq:PsiDef}) yields the full solution for the distribution function in the spatial-diffusion approximation,
\begin{equation} \label{eq:diffusion_solution}
\phi(z,\vect{\hat{q}}) =
\begin{cases}
 \phi_\infty & \mbox{for $z>0$ ;} \\
Ce^{\beta_u c z/\mathcal{D}_u}\left[1+\frac{\beta_u c^2}{\mathcal{D}_u}\MyPsiB(\vect{\hat{q}})\right] & \mbox{for $z<0$ ,}
\end{cases}
\end{equation}
where $C$ is a constant.
These expressions are valid far from the shock discontinuity, $\vert z\vert\gg \mathcal{D}/c$, where all parameters vary smoothly.
In order to check whether the approximation remains valid close to the shock, one must test if the continuity condition (\ref{eq:JumpDiff2}) can be satisfied identically (i.e., for any particle direction, $\vect{\hat{q}}$) at $z=0$.

Equation (\ref{eq:diffusion_solution}) implies that a consistent solution in the spatial-diffusion approximation requires that $\phi_{d1}=0$, indicating that the anisotropy must be second order downstream.
For such near isotropy, Eq.~(\ref{eq:transportKappa3}) is trivially satisfied, so there is no need to find a self-consistent solution to Eq.~(\ref{eq:psi}) downstream.
In the present framework, the result $\phi_{d1}=0$ holds, in particular, at the shock front, where Eq.~(\ref{eq:JumpDiff2}) reduces to
\beq
\phi_{u1}=-(2+p)(\beta_u-\beta_d)\phi_{u\iso}\mu \fin
\label{eq:JumpDiff3}\eeq
This relation therefore requires that $\MyPsiB_{u}\propto\mu$ at the shock front.
If this relation holds, then Eqs.~(\ref{eq:DDef}) and (\ref{eq:diffusion_solution}) yield $\phi_{u1}=-3\beta_u \phi_\iso \mu_u$. Using this result and the solution (\ref{eq:diffusion_solution}), Eq.~(\ref{eq:JumpDiff}) then becomes
\beq
C+C\left[(2+p)(\beta_u-\beta_d)-3\beta_u\right]\mu=\phi_{\infty} \fin
\eeq
This equation can be satisfied for any $\mu$ only if both $C=\phi_{\infty}$ and $p=(\beta_u+2\beta_d)/(\beta_u-\beta_d)$, which finally yields the standard spectrum (\ref{eq:DSAp}).

We reason that applying the approximation of spatial diffusion near the shock is self-consistent to first order in $\beta$, as necessary for the derivation of the spectrum, only if the upstream scattering function satisfies the necessary requirement $\MyPsiB_u\propto \mu$, or equivalently
\begin{equation} \label{Eq:ScndAnisoSpatDiff}
\int (\mu'-\mu)\MyScat_u(\vect{\hat{q}},\vect{\hat{q}}') d\Omega'\propto \mu \fin
\end{equation}
In such a case, it follows that $\MyPsi_{u1}(z=0)\propto\mu$, $\phi_{d1}(z)=0$, $C=\phi_\infty$, and so $p\simeq p_0$.

In the simple limit of an isotropic medium, Eq.~(\ref{Eq:ScndAnisoSpatDiff}) is satisfied, and the relation $\MyPsi_{u1}\propto\mu$ can indeed be seen to hold.
In such a medium, the scattering probability depends only on the angle between the initial and final particle directions, so we may write $\MyScat(\vect{\hat{q}},\vect{\hat{q}'})=\MyScat(\vect{\hat{q}\cdot \hat{q}}')$. Then Eq.~(\ref{eq:psi}) has an exact solution with the necessary form (\eg \citealt{PhysKinetics81}, section 11),
\beq
\MyPsiB=-\frac{\mu}{w_t} \coma
\eeq
where $w_t\equiv 4\pi\langle(1-\vect{\hat{q}\cdot\hat{q}}')\MyScat(\vect{\hat{q}\cdot \hat{q}}')\rangle$ is the transport scattering coefficient.
Here, the averaging is over $\vect{\hat{q}}'$, $w_t$ is independent of $\vect{\hat{q}}$, the diffusion coefficient is given by $\mathcal{D}=c^2/(3w_t)$, and the aforementioned implications, including $p=p_0+O(\beta)$, follow.

It is important to note, however, that the dependence $\MyPsi_1\propto\mu$ upstream is not universal.
In order to demonstrate that in an anisotropic medium, the angular distribution inferred in the spatial-diffusion approximation may be inconsistent with Eq.~(\ref{eq:psi}), consider the simple case of a separable scattering probability, $\MyScat(\vect{\hat{q}},\vect{\hat{q}}')=U(\vect{\hat{q}})U(\vect{\hat{q}}')$.
Here, the solution to Eq.~(\ref{eq:psi}) along with the normalization (\ref{eq:PhiOneNorm}) is
\beq
\MyPsiB(\mu)=\frac{1}{4\pi\langle U\rangle}
\left[\left\langle\frac{\mu}{U(\mu)}\right\rangle-\frac{\mu}{U(\mu)}\right] \coma
\eeq
which generally does not satisfy the continuity condition (\ref{eq:JumpDiff2}), and is thus inconsistent with a global spatial-diffusion approximation.
Equivalently, one can confirm that Eq.~(\ref{Eq:ScndAnisoSpatDiff}) is generally violated for a separable $\MyScat$.

The distribution function remains generally inconsistent with the requirement (\ref{Eq:ScndAnisoSpatDiff}) also in the small-angle scattering limit.
For instance, consider an upstream scattering function of the form
$\MyScat_u(\vect{\hat{q}},\vect{\hat{q}'})=h(\mu+\mu')\exp(-|a \,\vect{\hat{q}\cdot\hat{q}}'|)$,
where $a$ is a constant and $h$ is an arbitrary function.
This form is consistent with unitarity (\ref{eq:Unitarity}) and even with detailed balance, would be considered small-angle scattering for sufficiently large $|a|$, and becomes anisotropic for nontrivial $h$.
Testing Eq.~(\ref{Eq:ScndAnisoSpatDiff}), or equivalently evaluating the integral in Eq. (\ref{eq:psi}) by assuming $\psi\propto\mu$, is straightforward, especially for a strongly peaked $\MyScat$; for example, for $h(x)=x$ the integral is $\propto (3\mu^2-1)$.
Indeed, for a general choice of $h$, the integral is not $\propto \mu$, so deriving the spectrum by a global approximation of spatial diffusion is not self-consistent.

We see that in the general case, the consistency requirement (\ref{Eq:ScndAnisoSpatDiff}) is violated, even for small-angle scattering.
Here, non-diffusive effects at distances $|z|\lesssim \mathcal{D}/c$ from the shock front preclude a self-consistent derivation of the spectrum (\ref{eq:DSAp}) based on a global spatial-diffusion approximation.
Instead, one must solve the full transport equation (\ref{eq:transport_psi}), which in general can be done only numerically, as we discuss in \S\ref{sec:LargeAngleGen} and \S\ref{sec:DiffExamples}.
Moreover, we have assumed a sufficiently homogeneous or small anisotropy, $|\partial_z \phi_1|\ll |\partial_z\phi_\iso|$, to derive Eq.~(\ref{eq:transportKappa3}) and near isotropy (\ref{eq:PsiDef}) in diffusive regions, and assumed that the anisotropy remains small even at the shock front, to derive Eq.~(\ref{eq:JumpDiff}).
These assumptions cannot be a-priori justified, and indeed are not satisfied for all scattering functions.
We conclude that deriving the spectrum by globally invoking spatial diffusion is unwarranted for a general anisotropic medium, for which $p$ can in fact substantially deviate from $p_0$, as we later show.

The spectrum (\ref{eq:DSAp}) is nevertheless recovered in a medium which is not too anisotropic, although this was not yet proven, and should not be inferred from an inconsistent application of the spatial-diffusion approximation.
When, and how, does the spectrum deviate from Eq.~(\ref{eq:DSAp})?
We address this question for an arbitrary scattering function in \S\ref{sec:LargeAngle}, deferring the important limit of small-angle scattering to \S\ref{sec:SmallAngle}.

\section{Large-angle scattering}
\label{sec:LargeAngle}

In order to identify the circumstances under which the spectrum (\ref{eq:DSAp}) is valid, and quantify the deviations from this result, we consider here an arbitrary, large-angle scattering function. To frame the discussion, we begin with the simple case of isotropic, large-angle scattering, which to our knowledge was not rigorously solved until now.

\subsection{Isotropic scattering}
\label{sec:LargeAngleIso}

Consider the simple case in which the large-angle scattering is isotropic in the fluid frame.
We already know from \S\ref{sec:SpatialDiffAs} that in this case, the global approximation of spatial diffusion consistently yields the spectrum (\ref{eq:DSAp}).
It is nevertheless instructive to solve this specific problem without invoking spatial diffusion in the vicinity of the shock, but rather by matching asymptotic solutions at large optical depths from the shock.
The failure of this method for an arbitrary scattering function, as indicated in \S\ref{sec:LargeAngleGen}, demonstrates how spatial diffusion and Eq.~(\ref{eq:DSAp}) can break down for anisotropic scattering.

For isotropic scattering, the scattering rate is angle-independent, $\MyScat(z,q;\vect{\hat{q}},\vect{\hat{q}}')=\kappa_0(z,q)$, where we allowed for some inconsequential dependence on the distance from the shock and on momentum.
Here, the transport equation (\ref{eq:transport_psi}) becomes
\begin{equation}\label{eq:transportIso}
(\mu+\beta)\partial_\tau \phi(\tau,\mu) =
-\phi+\frac{1}{2}\int_{-1}^{1} \phi' \, d\mu' \coma
\end{equation}
where we defined the optical depth
\beq \label{eq:tau}
\tau \equiv \frac{4\pi}{c}\int_0^z \kappa_0(z',q) \,dz' \coma
\eeq
and, for brevity, $\phi'\equiv\phi(\tau,\mu')$.
It is useful to write the equation in the shock frame,
\begin{equation}\label{eq:transportSFrame}
\!\!\!\tilde{\mu}\partial_\tau \tilde{\phi}(\tau,\tilde{\mu}) = -\gamma^2(1-\beta\tilde{\mu})\tilde{\phi}+\frac{\int_{-1}^{+1}(1-\beta\mu')^{p}\tilde{\phi}'\,d\mu'}{2(1-\beta\tilde{\mu})^{p+1}} \fin
\end{equation}
As we focus here exclusively on the shock frame, the tilde that designates a shock-frame variable is omitted throughout the rest of \S\ref{sec:LargeAngleIso}.

Working in the non-relativistic shock limit, an expansion in powers of $\beta\ll1$ yields
\begin{eqnarray} \label{eq:transportApprox0}
\mu \partial_\tau \phi & = & \left[1+(p+1)\mu\beta+\frac{(p+2)(p+1)}{2}\mu^2\beta^2\right]N \nonumber \\
& & -(1-\mu\beta+\beta^2)\phi -\left[1+(p+1)\mu\beta\right]p\beta j \nonumber \\
& & + \frac{(p-1)p}{2}\beta^2\langle\mu^2\phi\rangle +O(\beta^3N) \coma
\end{eqnarray}
where we defined
\begin{equation} \label{eq:DefNandJ}
N\equiv\langle \phi \rangle \quad \mbox{and} \quad j\equiv\langle \mu\phi \rangle \coma
\end{equation}
in the shock frame, in accord with \S\ref{sec:NoSpatialDiff}.
Averaging this equation over $\mu$ then gives the moment equation
\begin{align} \label{eq:transportApprox1}
\frac{dj}{d\tau} = & \frac{(p+4)(p-1)}{6}\beta^2 N -(p-1)\beta j \\
& + \frac{(p-1)p}{2}\beta^2\langle\mu^2\phi\rangle +O(\beta^3N) \fin \nonumber
\end{align}

While the spatial diffusion approximation in general breaks down, as discussed in \S\ref{sec:SpatialDiffusion}, near the shock front $|\tau|\lesssim 1$, it does hold in the so-called diffusive regions, $|\tau|\gg1$, where the PDF is nearly isotropic.
Indeed, in the present problem, isotropy is guaranteed within a small number of scattering events, \ie within $\tau$ of order a few.
Hence, here it is unnecessary to invoke auxiliary assumptions regarding the confinement of streaming-like effects to the vicinity of the shock, as discussed in \S\ref{sec:NoSpatialDiff}.

In these diffusive regions, the anisotropic component $\Delta\phi\equiv \phi-N$ is small and slowly-varying, $|\Delta\phi|=O(\beta N)$ and $d(\Delta\phi)/d\tau=O(\beta N)$, so we may replace $\langle \mu^2\phi\rangle$ by $N/3$.
To first order in $\beta$, Eq.~(\ref{eq:transportApprox0}) then becomes
\begin{equation} \label{eq:transportApprox2}
\mu N'(\tau) = -\Delta\phi + (p+2)\mu\beta N +O(\beta^2 N)\coma
\end{equation}
such that Eq~(\ref{eq:transportApprox1}) yields
\begin{equation} \label{eq:transportApprox3}
N''(\tau)=3\beta N'(\tau) +O(\beta^3N) \fin
\end{equation}
Taking into account the boundary conditions, we conclude that
\begin{equation} \label{eq:SolDiff}
N(\tau\ll -1)=C_u e^{3\beta_u\tau} \smlspc \mbox{and} \smlspc
N(\tau\gg 1)=C_d \coma
\end{equation}
where $C_u$ and $C_d$ are constants, and at least $C_d$ must be of order unity.
Plugging these results into Eq.~(\ref{eq:transportApprox2}) and averaging over $\mu$ indicates that
\begin{equation} \label{eq:SolDiffFlux1}
j(\tau\ll -1)=\frac{p-1}{3}\beta_u C_u e^{3\beta_u\tau}
\eeq
and
\begin{equation} \label{eq:SolDiffFlux2}
j(\tau\gg 1)=\frac{p+2}{3}\beta_d C_d \fin
\end{equation}

The far upstream and far downstream solutions can now be matched, as follows.
In the diffusive regions, Eqs. (\ref{eq:SolDiff}--\ref{eq:SolDiffFlux2}) indicate that $j\sim \beta N$.
Equation (\ref{eq:transportApprox1}) then implies that $j$ varies over an optical depth scale $\tau\sim \beta^{-1}$.
Similarly, in the diffusive regions, Eq.~(\ref{eq:transportApprox0}) indicates that $N$ too varies over a scale $\tau\sim\beta^{-1}$. It is therefore possible to match both $N$ and $j$ across the shock, bridging the upstream and downstream diffusive regions.
More precisely, note that uniform $N$ and $j$, up to corrections of fractional order $\beta$, solve Eq.~(\ref{eq:transportApprox0}) in the non-diffusive, $|\tau|\lesssim $ a few, region, with Eqs.~(\ref{eq:SolDiff}--\ref{eq:SolDiffFlux2}) now used as boundary conditions.

Therefore, there exists a region $-\beta_u^{-1}\ll \tau\ll \beta_d^{-1}$, in which $N$ in the two diffusive region solutions of Eq.~(\ref{eq:SolDiff}) can be matched to leading order, so
\beq\label{eq:Matching1}
C_u=C_d + O(\beta) \coma
\eeq
as $C_d$ is of order unity.
Similarly, within this matching, $-\beta_u^{-1}\ll \tau\ll \beta_d^{-1}$ region, $j$ in the solutions (\ref{eq:SolDiffFlux1}) and (\ref{eq:SolDiffFlux2}) can be matched to leading order, yielding
\beq\label{eq:Matching2}
(p-1)\beta_u C_u=(p+2)\beta_d C_d + O(\beta^2) \fin
\eeq
Combining Eqs.~(\ref{eq:Matching1}) and (\ref{eq:Matching2}) now implies the spectrum in Eq.~(\ref{eq:DSAp}).

\subsection{Large-angle scattering: modified spectra}
\label{sec:LargeAngleGen}

Consider the generalization of \S\ref{sec:LargeAngleIso} for an arbitrary, large-angle scattering function,  $\MyScat(z,q;\vect{\hat{q}},\vect{\hat{q}}')$.
It is clear from \S\ref{sec:SpatialDiffAs} that for such general $\MyScat$, applying the spatial diffusion approximation near the shock front can lead to an inconsistency, suggesting that substantial deviations from the spectrum (\ref{eq:DSAp}) may be possible.
Here, we quantify the conditions and implication of such a spectral deviation, and demonstrate that it can indeed manifest.

As long as angular variations in $\MyScat$ are sufficiently small, the problem can be solved in methods similar to those indicated above, in particular the asymptotic matching method used in \S\ref{sec:LargeAngleIso}.
This method works as long as there exists a well-defined matching region around the shock front, which simultaneously shows a well-isotropized PDF and unmodified $N$ and $j$.
The problem becomes difficult when $\MyScat$ spans a wide, typically $\gtrsim \beta^{-1}$ range of values as a function of angles, at a given $z$ and $q$.
For simplicity, the dependence of $\MyScat$ upon $z$ and $q$ is assumed separable, and is henceforth omitted.

One difficulty is that in such cases, there is no obvious way to define an optical depth $\tau$ independent of angle $\vect{\hat{q}}$, in analogy with Eq.~(\ref{eq:tau}).
Other difficulties arise if the PDF does not isotropize sufficiently quickly in the putative matching region, allowing for leading order variations in $N$ or $j$;
such effects may be considered in part as manifestations of the streaming problem outlined in \S\ref{sec:NoSpatialDiff}.
Due to these difficulties, highly anisotropic scattering can diminish the matching region, causing it to disappear entirely, or to become misaligned for different angles $\vect{\hat{q}}$, rendering matching impossible.

Let us relate the spectrum to the global properties of the PDF, without invoking the spatial diffusion approximation near the shock.
Using the parametrization (\ref{eq:diffus}--\ref{eq:PhiOneNorm}), Eq.~(\ref{eq:meanTrans}) becomes
\begin{equation} \label{eq:meanTrans3}
\beta \phi_\iso + \langle \mu \phi_1 \rangle =
\begin{cases}
\beta_d \phi_\infty & \mbox{for $\tau>0$ \,;} \\
0 & \mbox{for $\tau<0$ \coma}
\end{cases}
\end{equation}
where $\phi$ is measured again in the fluid frame (henceforth).
The continuity relations (\ref{eq:JumpDiff0}) and (\ref{eq:JumpDiff2}), with their underlying assumption $|\phi_1|\ll\phi_0$, now relate the spectrum to the PDF behavior downstream,
\begin{equation} \label{eq:pFromAnisotropy}
\frac{p -1}{p_0-1} \simeq 1+\frac{\langle \mu \phi_{d1\sh} \rangle}{\beta_d} = 1+\frac{\langle \mu \phi_d(z=0) \rangle}{\beta_d}
\end{equation}
and
\begin{equation}\label{eq:pFromEvolution}
\frac{p -1}{p_0-1} \simeq \frac{\phi_\infty}{\phi_{\iso\sh}} =  \frac{\langle \phi(z\to\infty)\rangle}{\langle\phi(z=0)\rangle} \coma
\end{equation}
where subscript $s$ designates the shock front.
Thus, the spectrum is directly related both to the anisotropy $\phi_1$ at the shock front, and to the overall evolution of $\langle \phi\rangle$ downstream.
In particular, the spectrum (\ref{eq:DSAp}) requires the anisotropy measure $\langle \mu \phi_{ds} \rangle$ to be $O(\beta^2)$, and the evolution parameter $-1+{\phi_\infty}/{\phi_{\iso\sh}}$ to be $O(\beta^1)$.

Next, let us rewrite the transport equation as a diffusion-like equation.
Define some characteristic scattering rate $\kappa_0$, and a corresponding optical depth $\tau$ as in Eq.~(\ref{eq:tau}), chosen for example such that the normalized scattering function $\MyScatB(\vect{\hat{q}},\vect{\hat{q}}')\equiv \MyScat/\kappa_0$ averaged over $\Omega$ and $\Omega'$ is unity.
The transport equation (\ref{eq:transportKappa2}) can now be written in the form
\begin{equation}\label{eq:transport3}
(\mu+\beta)\partial_\tau(\phi_\iso+\phi_1) = \int \MyScatB(\vect{\hat{q}},\vect{\hat{q}}')\phi_1(\vect{\hat{q}}')\,d\Omega'-W\phi_1 \coma
\end{equation}
where we defined $W(\vect{\hat{q}})\equiv \int \MyScatB(\vect{\hat{q}},\vect{\hat{q}}')\,d\Omega'$.
Multiplying Eq.~(\ref{eq:transport3}) by $\mu/W$, averaging over $\Omega$, and rearranging, we arrive at
\begin{eqnarray} \label{eq:transport4}
\langle{\mu\phi}\rangle & = & -\partial_\tau \langle \mathcal{D}\phi \rangle  + \Pi_{\st} \coma \\
& = & -\langle \mathcal{D}\rangle \partial_\tau \phi_\iso + \Pi_{\st} -\partial_\tau \langle \mathcal{D} \phi_1 \rangle \coma \nonumber
\end{eqnarray}
where we defined a spatial diffusion function
\beq
\mathcal{D}(\vect{\hat{q}})\equiv \frac{(\mu+\beta)\mu}{W} \coma
\eeq
and a streaming term
\beq
\Pi_{\st}\equiv \left\langle \frac{\mu}{W} \int \MyScatB(\vect{\hat{q}},\vect{\hat{q}}') \phi_1(\vect{\hat{q}}')d\Omega' \right\rangle \fin
\eeq

The transport equation thus reduces to a spatial diffusion equation, with two important modifications to the flux: a streaming term $\Pi_{\st}$, as anticipated in \S\ref{sec:NoSpatialDiff}, and an anisotropy-dependent diffusive term $\partial_\tau \langle \mathcal{D} \phi_1 \rangle$.
Both of these terms can be neglected in the limit of isotropic scattering, in which case the integral in $\Pi_{\st}$ vanishes, and the only term surviving in $\partial_\tau \langle \mathcal{D} \phi_1 \rangle$ is small, of order $\beta\phi_1$.
When scattering is anisotropic but sufficiently close to being isotropic, these two terms can still be approximately neglected, so the spectrum (\ref{eq:DSAp}) is recovered, even though its derivation in \S\ref{sec:SpatialDiffAs} becomes inconsistent once the condition (\ref{Eq:ScndAnisoSpatDiff}) is violated.
In the more general case, however, neither term can be neglected, and the spectrum can become modified.

Figure \ref{fig:LA} demonstrates substantial deviations from the spectrum (\ref{eq:DSAp}), obtained with highly anisotropic, large-angle scattering, for arbitrarily small $\beta$.
In order to establish such a behavior, we choose scattering functions that yield a simple behavior $p=p(\beta)\neq p_0$ of the spectral index.
One such choice is $\MyScat_1=\exp[\alpha(\mu+\mu')]$, with the same constant $\alpha$ used upstream and downstream, shown in the figure as filled symbols (with a solid line to guide the eye).
As the figure shows, $p$ in this case is approximately a function of $\beta^\xi \alpha$, where $\xi\simeq 0.75$.
Assuming that such a scaling persists for arbitrarily small $\beta$, one can always find a sufficiently negative $\alpha$ that yields a noticeable deviation from $p_0$.

\begin{figure}[h!]
    \centerline{\hspace{-0.cm}\epsfxsize=8.7cm\epsfbox{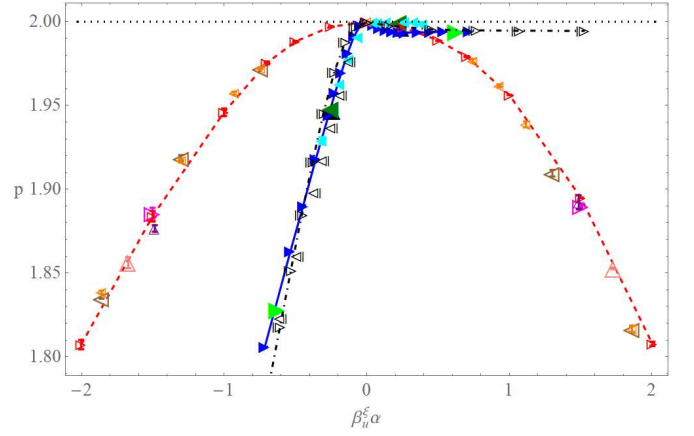}}
	\caption{\label{fig:LA}
    Substantial spectral deviation from the standard spectrum, $p_0=2$ (dotted horizontal line), due to anisotropic scattering in a non-relativistic shock of compression ratio $r=4$.
    For simplicity, the same scattering function is used upstream and downstream.
    Results are shown (symbols with numerical error bars and lines to guide the eye) for large-angle scattering functions $\MyScat_1(\vect{\hat{q}},\vect{\hat{q}}')=\exp[\alpha(\mu+\mu')]$ (filled blue-to-green symbols, solid line, scaled with $\xi=0.75$) and $\MyScat_2=\exp[\alpha(\mu^2+\mu'^2)]$ (empty red-to-orange symbols, dashed line, $\xi=0.43$),
    and for an angular diffusion function $D(\mu)=\exp(\alpha\mu)$ (black bar-triangles, dot-dashed; $\xi=0.8$).
    For such scattering, $p$ is approximately a function of $\beta^\xi\alpha$ (abscissa), so different shocks (right triangles for $\beta_u=0.04$, up triangles for $0.02$, and left triangles for $=0.004$) yield overlapping curves.
    For sufficiently small $\beta$, $p$ is insensitive to the frame in which $\mu$ and $\mu'$ are measured (large symbols for fluid frame, small for shock frame).
    	}
\end{figure}

The figure pertains to a compression ratio $r=4$, corresponding to a strong shock in a gas of adiabatic index $5/3$, such that $p_0=2$.
The results are obtained in the method of angular moments \citep{Keshet06}, generalized (Arad et al. 2020, in preparation) for large-angle scattering.
Numerical convergence is demonstrated by error bars, typically too small to see in the figure.

A second, simple choice of scattering function is $\MyScat_2=\exp[\alpha(\mu^2+\mu'^2)]$, shown in the figure as empty symbols (with a dashed line).
Here, $p$ is approximately a function of $\beta^\xi \alpha$ with $\xi\simeq 0.43$, and substantially deviates from $p_0$ for both positive and negative $\alpha$.
This scattering function satisfies the symmetries $\MyScat_2(\mu,\mu')=\MyScat_2(-\mu,\mu')$ and $\MyScat_2(\mu,\mu')=\MyScat_2(-\mu',-\mu)$.
The former symmetry guarantees that the streaming term $\Pi_\st$ vanishes in this case, so the spectral deviation here may be attributed to the $\partial_\tau \langle \mathcal{D} \phi_1 \rangle$ term.

These simple examples, while useful for demonstrative purposes, involve substantial variations in $\kappa$ over its domain.
These variations are exponential in $\beta^{-1}$, and therefore  become non-physically large in the small $\beta$ limit.
However, one can identify scattering functions which are only modestly anisotropic, and yet produce substantial deviations in the spectrum.
It is more convenient to analyze such scattering functions in the small-angle scattering limit, which we now turn to.

\section{Small-angle scattering}
\label{sec:SmallAngle}

Next, consider the small-angle scattering limit, also known as the limit of angular diffusion or pitch angle diffusion, typically assumed to provide a good approximation for astronomical shocks.
This is a special case of the arbitrary scattering discussed in \S\ref{sec:LargeAngleGen}, so constraints on the spectrum such as Eqs.~(\ref{eq:pFromAnisotropy}) and (\ref{eq:pFromEvolution}) apply.
In particular, as established in \S\ref{sec:SpatialDiffAs}, the standard spectrum $p=p_0$ is guaranteed for an isotropic medium in this limit, too.
It is not a priori clear, however, if substantial deviations from $p_0$, as demonstrated in \S\ref{sec:LargeAngleGen}, are possible for anisotropic small-angle scattering.
Let us start by analyzing the anisotropy pattern at the shock front, and showing that previous claims that first-order anisotropies cannot affect the spectrum are incorrect.

\subsection{Shock front anisotropy}
\label{sec:Anisotropy}

Our analysis is carried out in the fluid frame.
For diffusion in the direction $\vect{\hat{q}}$ of fluid frame momentum $\vect{q}$, one typically invokes detailed balance and axial symmetry, so the stationary transport equation becomes \citep{KirkSchneider87}
\beq
(\beta_i+\mu_i)c\gamma_i \frac{\partial f_i}{\partial z}  = \frac{\partial}{\partial \mu_i} \left[ (1-\mu_i^2) \bar{D}_i\frac{\partial f_i}{\partial \mu_i}\right]
\fin \label{eq:transport1} \eeq
The angular diffusion function, $\bar{D}=\bar{D}(z,q,\mu)$, introduces a length scale $c/\bar{D}$, which can be absorbed by rescaling $z$. In particular, one often assumes that $\bar{D}$ is separable, for example in the form $\bar{D}=D(\mu) D_2(z,q)$, where $D(\mu)$ is dimensionless and of order unity.
While this assumption is not essential for our main result, we adopt it here for convenience.
With the parametrization (\ref{eq:PhiDef}), the transport equation (\ref{eq:transport1}) then becomes
\beq (\beta+\mu) \partial_\tau \phi = \partial_\mu [(1-\mu^2) D(\mu) \partial_\mu \phi] \coma \label{eq:transport2} \eeq
where $\tau \equiv (\gamma c)^{-1} \int_0^z D_2(z',q) dz'$ is the optical depth, and the frame index $i$ is again omitted when possible.

The solution of Eq.~(\ref{eq:transport2}) under the boundary conditions (\ref{eq:boundary}) fixes the spectral index $p$.
Such derivations of the spectrum were carried out for an arbitrarily relativistic shock, numerically, for example by expanding $\phi$ in eigenfunctions in the downstream \citep{KirkSchneider87,HeavensDrury88} or in the upstream \citep{Kirk_2000}, or by a relaxation code \citep{NagarKeshet19}; semi-analytically, by evolving the moments of $\phi$ \citep{Keshet06}; and analytically \citep{Keshet_2005}, by approximating the downstream anisotropy at the shock-grazing angle, $\mu= -\beta$, as fixed by its non-relativistic shock limit.

For the present study, it is useful to reproduce some results from the \citet{Keshet_2005} shock-grazing analysis.
We assume an analytic behavior at the grazing angle, $\mu=-\beta$.
Expanding $\phi$ and $D$ in the fluid frame around this angle,
\beq \label{eq:expansion} \phi(\mu,\tau) = a_0(\tau) + a_1(\tau)(\mu+\beta) + a_2(\tau)(\mu+\beta)^2 + \cdots  \coma
\eeq
and
\beq\label{eq:Dexpansion}
D(\mu) = 1 +  d_1(\mu+\beta) + d_2(\mu+\beta)^2+\cdots \coma
\eeq
the transport equation (\ref{eq:transport2}) implies that for any $\tau$,
\beq \label{eq:coeff} a_2(\tau)=-\gamma^2 a_1(\tau)\left(\beta + d/2\right) \fin  \eeq
Here, $d\equiv \gamma^{-2}d_1$ is a measure of the deviation from isotropic diffusion, and for simplicity we chose the $\tau$ scaling such that $D(\mu=-\beta)=1$.
Continuity then yields an exact relation between the spectrum and the low-order expansion coefficients in Eq.~(\ref{eq:expansion}--\ref{eq:Dexpansion}), which can be used to constrain the spectrum, test numerical simulations, and give an accurate expression for the spectral index for arbitrary $\beta_u$ and $\beta_d$ \citep{Keshet_2005}.

In the non-relativistic shock limit, $\beta\ll1$, this spectrum--anisotropy relation gives \citep{Keshet_2005}
\begin{align}\label{eq:GrazingAniso}
\frac{a_1}{a_0} \, & \MyEqual_d
\gamma_d^2 (p+2) \frac{(p+2)(\beta_u-\beta_d)-(\beta_u+\beta_d+d_u)}{2(p+2)-(d_u-d_d)/(\beta_u-\beta_d)} \nonumber \\
\, & = \, \beta_u  \frac{\beta_u-\beta_d/2-d_u/2}{\beta_u+(d_d-d_u)/6} + O(\beta^2)
\end{align}
in the downstream frame; an analogous relation is obtained upstream.
Here, and henceforth, a subscript $d$ (subscript $u$) on the equality sign indicates that the calculation is carried out at the shock front in the downstream (upstream) frame, and variables without a fluid-frame index $i$ should be treated as such.
In the second equality in Eq.~(\ref{eq:GrazingAniso}), we have assumed that the classical spectrum (\ref{eq:DSAp}) is valid and that the diffusion is nearly isotropic on both sides of the shock, $d_u=d_d=0$.

Equation (\ref{eq:GrazingAniso}) shows that the shock-front anisotropy has derivatives that are first-order in $\beta$ in the downstream frame; a similar conclusion is reached in the upstream and shock frames.
Indeed, for isotropic angular diffusion, $d=0$, the downstream relation (\ref{eq:GrazingAniso}) reduces to
\beq \label{eq:GrazingIso}
\frac{a_1}{a_0} \, \MyEqual_d \,
\frac{p+1}{2}\gamma_d^2\beta_u-\frac{p+3}{2}\gamma_d^2\beta_d
\, = \, \beta_u-\frac{\beta_d}{2}+O(\beta^3)\coma
\eeq
where we again assumed that $p=p_0$ in the last equality.

Note that while the normalized derivative at $\mu=-\beta$, namely $(\phi'/\phi)_{\mu=-\beta}=a_1/a_0$,
is first order in $\beta$, the variations in $\phi(\mu)$ over the domain $-1\leq\mu\leq 1$ can in principle be smaller, only of second order.
Such a small, $O(\beta^2)$ anisotropy can be reconciled with order $\beta$ angular derivatives only if the latter are confined to a narrow, $O(\beta)$ range of $\mu$.
This peculiar behavior can indeed emerge in the downstream frame, albeit generally not in the upstream or shock frames, as we verify in the case of isotropic angular diffusion.
Regardless, the following discussion holds quite generally for any $O(\beta)$ anisotropies.

Next, we address the problem of deriving the spectrum in the small-angle scattering, non-relativistic shock, limit.
Analytically solving the transport equation (\ref{eq:transport2}) has proved to be difficult even in this limit, so we resort to an alternative method for computing $p$.

\subsection{Spectral sensitivity to $O(\beta_{\Myd})$ anisotropies}
\label{sec:AnisotropyImportant}

A useful approach \citep{Fermi49, Bell_1978} for computing the spectrum is to relate the spectral index to the fractional energy-gain $\Gain$ in a Fermi cycle, and the probability $\myPesc$ that a particle crossing the shock downstream escapes and never returns upstream,
\beq p \simeq 1 - \frac{\ln(1-\myPesc)}{\ln \MeanGainP} \fin \label{eq:SvsPret} \eeq
Here, $\llangle \ldots\rrangle$ designates flux-averaging, defined explicitly below.
Both $\myPesc$ and $\Gain$ can be computed, at least approximately, if the angular PDF at the shock front, $\phi_{\sh}(\mu)\equiv \phi(\tau=0,\mu)$, is known.

Working in the downstream frame, consider a particle undergoing a Fermi cycle in which it crosses the shock from the downstream to the upstream at some angle $\mu_{-}$ satisfying $-1<\mu_{-}<-\beta$, and returns to the downstream at an angle $-\beta<\mu_{+}<1$.
The energy gain during such a cycle is
\beq \label{eq:CycleGain}
1+\Gain \MyEqual_d
{(1-\beta_r \mu_-)}/{(1-\beta_r \mu_+)}\fin
\eeq
The mean energy gain may be computed by averaging Eq.~(\ref{eq:CycleGain}) over the shock-frame flux across the shock front,
\beq
dj \equiv (\mu+\beta)\phi(\mu)d\mu \coma
\eeq
in all $\mu_-$ and $\mu_+$ directions,
\begin{equation} \label{eq:GainD}
\MeanGainP \MyEqual_d \left\llangle \frac{1-\beta_r \mu_{-}}{1-\beta_r\mu_{+}}\right\rrangle \MyEquiv_d \frac{\mathlarger{\int} \frac{1-\beta_r\mu_{-}}{1-\beta_r\mu_{+}}dj^{-}dj^{+}}{\int dj^{-} dj^{+}} \fin
\end{equation}
Here, an integral over $dj^\pm$ is taken over the full range in which $\mu_\pm$ is defined.
An analogous expression is obtained in the upstream frame,
\begin{equation} \label{eq:GainU}
\MeanGainP  \MyEqual_u \left\llangle\frac{1+\beta_r \mu_{+}}{1+\beta_r\mu_{-}}\right\rrangle \MyEquiv_u \frac{\mathlarger{\int} \frac{1+\beta_r\mu_{+}}{1+\beta_r\mu_{-}}dj^{-}dj^{+}}{\int dj^{-} dj^{+}} \fin
\end{equation}
Note that the shock-frame flux $j=\int dj$ is consistent with its usage in \S\ref{sec:SpatialDiffusion} and \S\ref{sec:LargeAngle}, and with definition (\ref{eq:DefNandJ}), in which $\phi$ is measured in the shock frame.

If the correlations between $\mu_{-}$ and $\mu_{+}$ can be neglected, and the shock front PDF $\phi_{\sh}(\mu)$ is known, the integrals in Eqs.~(\ref{eq:GainD}) or (\ref{eq:GainU}) can be carried out.
These correlations are indeed negligible in the non-relativistic shock regime, where the PDF becomes nearly isotropic, as we confirm numerically.
Moreover, correlations are found (Arad et al., in preparation) to be negligible even for relativistic shocks, provided that $\MeanGain$ is computed in the upstream frame, \ie using Eq.~(\ref{eq:GainU}) rather than (\ref{eq:GainD}).
These conclusions apply even for highly anisotropic scattering functions, shown in \S\ref{sec:DiffExamples} to yield substantial deviations from the spectrum (\ref{eq:DSAp}).

Motivated by Eqs.~(\ref{eq:coeff}--\ref{eq:GrazingIso}),
we henceforth take into account small shock-front anisotropies, in the form
\beq \label{eq:PhiZero}
\phi_{\sh}(\mu) \MyEqual_d \phi_0+\phi_1(\mu) \MyEqual_d 1 + \beta \, \delta \phi(\mu) \coma
\eeq
where the normalization of the leading term is arbitrarily chosen as unity.
We have parametrized $\phi_1=\beta\,\delta \phi$ in anticipation of the typical case, where anisotropy is first order and so $\delta \phi=O(\beta^0)$, although anisotropies of both second order (for isotropic scattering, see \S\ref{sec:SpatialDiffAs}) or zeroth order (see \S\ref{sec:DiffExamples}) are possible.

The mean energy gain is then found from Eq.~(\ref{eq:GainD}) to be \citep{Bell_1978}
\beq \label{eq:Gain}
\MeanGain = \frac{4}{3}(\beta_u-\beta_d) + O(\beta^2) \coma
\eeq
independent of $\delta \phi$.
One sees that $O(\beta)$ anisotropies have no leading-order effects on the mean gain, as long as correlations can be neglected.
Moreover, we have not identified any departure from Eq.~(\ref{eq:Gain}) in a non-relativistic shock, even for highly anisotropic diffusion functions that yield an order unity PDF anisotropy and a substantial deviation from the spectrum (\ref{eq:DSAp}).

This is not the case, however, for the escape probability.
Considering the particle flux crossing toward the downstream, $j^+=\int dj^+>0$, and the fraction of this flux returning upstream, $j^-=\int dj^-<0$, the escape probability can be written as
\begin{equation}
\label{eq:Pret}
\myPesc
\, \MyEqual_d \, 1+\frac{j^-}{j^+} \, \MyEqual_d \, \frac{j}{j^+} \fin
\end{equation}
Here, the total flux $j\equiv j^++j^-$ is constant on each side of the shock, as seen by integrating Eq.~(\ref{eq:transport2}); in particular, in the downstream $j(\tau>0)=2\beta_d \phi_\infty$.
Unlike the energy gain, the escape probability is independent of Fermi-cycle correlations, and so can be computed exactly if $\phi_{\sh}$ is precisely known, without additional assumptions.

In particular, the escape probability can be computed from the first equality in Eq.~(\ref{eq:Pret}) if the PDF were to be assumed isotropic,
\begin{equation}
\label{eq:PretIso}
\myPescIso = 4\beta_d+O(\beta^2) \fin
\end{equation}
In this isotropic limit, $\myPescIso$ can be computed also in alternative methods, by invoking random-walk or spatial diffusion arguments, as discussed above.
If the isotropy assumption could be justified, Eqs.~(\ref{eq:SvsPret}), (\ref{eq:Gain}), and (\ref{eq:PretIso}) would directly lead \citep{Bell_1978} to the well-known spectrum of Eq.~(\ref{eq:DSAp}).

It is important to note, however, that even the leading-order term in $\myPesc$ is sensitive to $O(\beta^1)$ anisotropies in the shock-front PDF of Eq.~(\ref{eq:PhiZero}), measured in the downstream frame.
Indeed, the first equality in Eq.~(\ref{eq:Pret}) indicates that
\beq \label{eq:PescAniso}
\myPesc \MyEqual_d 4\beta(1+I/2)+O(\beta^2) \coma
\eeq
where
\beq \label{eq:IDef}
I \MyEquiv_d \frac{\langle \mu\phi \rangle}{\beta/2} \MyEquiv_d \beta^{-1}\int_{-1}^1 \mu\, \phi_{\sh} \,d\mu \MyEqual_d \int_{-1}^1 \mu\, \delta\phi \,d\mu
\eeq
is the shock-front flux in the downstream frame.
Thus, the escape probability (\ref{eq:PretIso}), computed for an isotropic PDF, can only be used if the first-order anisotropy integral $I$ vanishes.
Otherwise, combining Eqs.~(\ref{eq:SvsPret}), (\ref{eq:Gain}) and (\ref{eq:PescAniso}) yields a spectral index $p=(r+2+3I/2)/(r-1)+O(\beta)$, that deviates to leading order from Eq.~(\ref{eq:DSAp}).
This result is equivalent to Eq.~(\ref{eq:pFromAnisotropy}), and can be directly obtained by adopting Eq.~(\ref{eq:PhiZero}) and requiring that $j_u=0$.

As a concrete example, consider using only one or two of the lowest order terms in the expansion (\ref{eq:expansion}), inferred from Eq.~(\ref{eq:DSAp}) for the simple case of isotropic diffusion; namely, using $a_1$ from Eq.~(\ref{eq:GrazingIso}), with or without $a_2$ from  Eq.~(\ref{eq:coeff}), and neglecting higher-order terms.
The escape probability derived in this case from Eq.~(\ref{eq:Pret}) is $\myPesc = (4\beta_u+10\beta_d)/3+O(\beta^2)$,
which leads in the non-relativistic shock limit to the spectral index $p=(2\beta_u+3\beta_d/2)/(\beta_u-\beta_d)=(2r+3/2)/(r-1)$, inconsistent with Eq.~(\ref{eq:DSAp}). In particular, this expression yields $p=19/6\simeq 3.17$ (instead of the classical $p_0=2$) in the limit of a strong shock in a medium with $\Gamma=5/3$.

\subsection{Conditions for spectral modification}
\label{sec:pDev}

One can also relate the spectrum to the spatial evolution of the PDF downstream.
To leading order, the second equality in Eq.~(\ref{eq:Pret}) and the parametrization (\ref{eq:PhiZero}) yield $\myPesc =4\beta_d \phi_\infty/1$.
This links the escape probability --- and thus also the spectrum and the shock-front flux $I$ --- to the evolution in $\phi_{\iso}$ between the shock and the far downstream.
In particular,
\begin{equation}
\left[\phi_{\iso}\right]_{\tau=0}^\infty = \phi_\infty - 1  \MyEqual_d \frac{1}{2}I+O(\beta) \fin
\end{equation}
Similarly, Eq.~(\ref{eq:Pret}) relates the escape probability to the evolution in the forward flux, $j^+(\tau)$, between the shock front and far downstream, $\myPesc\MyEqual_d 4\beta j^+_{\infty}/j^+_{\sh}$.

We see that in the downstream frame, the deviation of the spectrum from Eq.~(\ref{eq:DSAp}) is directly related to the first-order anisotropy of the shock-front PDF, and to the zeroth order spatial variations in the PDF and in the forward flux.
To leading order in $\beta$, these relations become
\begin{equation}\label{eq:DeltaP}
\frac{p - 1}{p_0-1}
\MySimeq_d 1+\frac{I}{2}
\MySimeq_d \phi_\infty
\MySimeq_d \frac{j^+_{\infty}}{j^+_{\sh}}
\fin
\end{equation}
These results conform with, and supplement, relations (\ref{eq:pFromAnisotropy}) and (\ref{eq:pFromEvolution}).

Figure \ref{fig:DiffusionExample} demonstrates the downstream PDF and its relation to the spectrum for the case of mildly anisotropic diffusion, $D_u=1+0.4\mu$ and $D_d=1-0.4\mu$, for which the spectrum $p=p_0+O(\beta)$ remains consistent with Eq.~(\ref{eq:DSAp}).
For simplicity, we choose a shock with a modest compression ratio, $r=5/4$, to clearly distinguish between different powers of $\beta$.
The PDF is computed in three different methods: an expansion in downstream eigenfunctions \citep{KirkSchneider87}, an expansion in upstream eigenfunctions \citep{Kirk_2000}, and a moment expansion \citep{Keshet06}.
The three methods are seen to give consistent results; the latter converges faster and avoids spurious oscillations near the $\mu=\pm1$ poles, allowing one to reach smaller values of $\beta$.

\begin{figure}[h!]
    \hspace{-0.2cm}
    \centerline{\epsfxsize=9.1cm
    \epsfbox{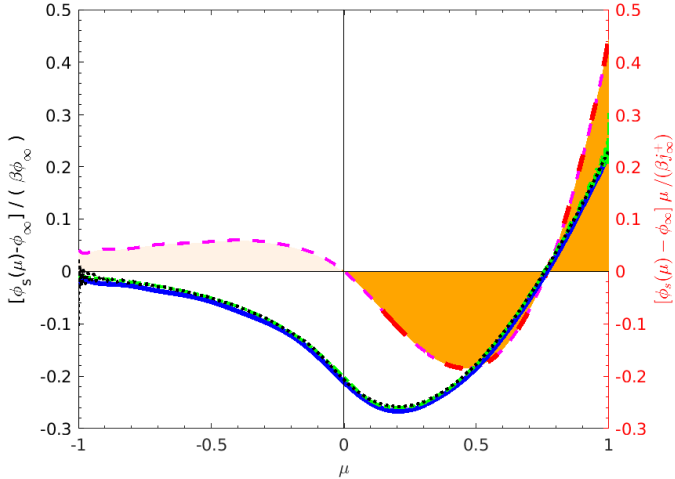}}
	\caption{\label{fig:DiffusionExample}
Downstream-frame analysis of the angular PDF at an $r=5/4$ shock front with anisotropic angular diffusion, $D_u=1+0.4{\mu}$ and $D_d=1-0.4{\mu}$.
The non-uniform component of the PDF, $\phi-\phi_\infty$, is shown (left axis) normalized by $\beta\phi_\infty$, computed in the moment method for $\beta_u=0.001$ (using 10 Legendre moments; solid blue curve) and in the eigenfunction method for $\beta_u=0.01$, using $n=32$ upstream (dotted black) or downstream (dot-dashed green) functions.
The spectrum is given by Eq.~(\ref{eq:DSAp}), as one can infer from the $O(\beta^2)$ flux and $O(\beta^1)$ evolution downstream.
Namely, Eq.~(\ref{eq:DeltaP}) guarantees that $p=p_0+O(\beta)$ in several ways: (i) $I/j^+_{\infty}$ (total shaded region inside short-dashed magenta; right axis) is only of order $O(\beta^1)$, (ii) $(-1+j^+_{\sh}/j^+_{\infty})/\beta$ (dark shaded region inside long-dashed red) is only of order $O(\beta^0)$, and (iii) $(\phi_{\sh}-\phi_\infty)/\phi_\infty$ is only of order $O(\beta^1)$.
	}
\end{figure}

By examining $\phi_{\sh}(\mu)-\phi_\infty=_d\phi_{\sh}(\mu)-j/(2\beta)$, one can infer from the figure both the angular and the spatial properties of the downstream PDF, and confirm that they are all consistent with the spectrum (\ref{eq:DSAp}).
First note that while the anisotropy of $\phi_{\sh}(\mu)$ is first order in $\beta$, its $\mu$-weighted integral (\ref{eq:IDef}) gives only a second order result (full shaded region), such that $I=O(\beta)$.
Next, note that $\phi_{\sh}(\mu)-\phi_\infty$ too is first order in $\beta$, indicating no zeroth-order evolution in $\phi_\iso$ between the shock and far downstream, such that $\phi_\infty-1=O(\beta)$.
We deduce that the $(\mu+\beta)$ weighted integral (dark shaded region) of $\phi_{\sh}(\mu)-\phi_\infty$ is also of order $\beta$, and so $j^+_\infty-j^+_{\sh}=O(\beta)$.
These conclusions provide three different perspectives on the result $p=p_0+O(\beta)$, as summarized by Eq.~(\ref{eq:DeltaP}).
Conversely, a substantial modification to the spectrum (\ref{eq:DSAp}) would necessitate an order $\beta^0$ contribution in all three parameters: $I$, $\phi_{\sh}(\mu)-\phi_\infty$, and $j^+_\infty-j^+_{\sh}$.

As the anisotropy is typically small, the isotropic component $\phi_\iso$ of the PDF evolves slowly.
Formally, the first term in the expansion of the transport equation (\ref{eq:transport2}) in powers of $\beta$ gives $\pr_\tau \phi_\iso=0$, seemingly implying a uniform isotropic component \citep[\eg][]{Blandford_Eichler_87}.
However, $\phi_\iso$ evolves over a scale $\tau\gtrsim\beta^{-1}$, and so is approximately uniform only if the PDF isotropizes within an optical depth of order $\beta^0$ from the shock. (Indeed, the same formal argument could be invoked to argue for a constant $\phi_\iso$ upstream, and would necessarily fail, for the same reason.)
Such nearby isotropization would indeed render $\phi_{\sh}-\phi_\infty=O(\beta)$, guaranteeing the spectrum (\ref{eq:DSAp}).
To see this, integrate Eq.~(\ref{eq:transport2}) over $0<\tau<\infty$, giving
\begin{align} \label{eq:DownstreamEv}
\frac{p-p_0}{p_0-1} & \simeq  \phi_\infty-1 \\
& \MyEqual_d  \beta \,\delta\phi_{\sh} +\frac{\beta}{\mu+\beta}\pr_\mu \left[ (1-\mu^2) D(\mu) \pr_\mu \int_{0}^{\infty} \delta\phi \, d\tau \right] \fin \nonumber
\end{align}
As long as the angular diffusion is not far from being isotropic, $\delta\phi$ decays, roughly exponentially, within $\tau\sim$ a few.
This can be inferred, for example, from the smallest downstream eigenvalue of the transport equation being of order unity.
For approximately isotropic diffusion, the RHS of Eq.~(\ref{eq:DownstreamEv}) is therefore of order $\beta$, so Eq.~(\ref{eq:DeltaP}) guarantees that $p\simeq p_0$.

Conversely, a substantial modification to the spectrum (\ref{eq:DSAp}) would necessitate a $\sim \beta^{-1}$ contribution to the integral in Eq.~(\ref{eq:DownstreamEv}), corresponding to an anisotropy persisting at distances $\tau\gtrsim\beta^{-1}$ downstream of the shock.
Furthermore, such a substantial anisotropy must also be present at the grazing angle, $\mu\simeq -\beta$, because integrating Eq.~(\ref{eq:transport2}) over $-\beta<\mu<1$ yields $\partial_\tau j^+=-a_1/\gamma^{2}$.
Assuming an analytic grazing-angle behavior, either Eq.~(\ref{eq:GrazingAniso}) or Eq.~(\ref{eq:GrazingIso}) imply that $a_1=O(\beta)$.
Hence, a substantial contribution to $j^+_{\sh}-j^+_\infty$, and therefore to $p-p_0$, would require the grazing anisotropy $a_1$ to remain of order $\beta$ even at distances $\tau\sim \beta^{-1}$.

\subsection{Spectrum-modifying diffusion functions}
\label{sec:DiffExamples}

In summary, a noticeable, order unity deviation from the classical spectrum (\ref{eq:DSAp}) would require the PDF anisotropy $\phi_1=\beta\,\delta \phi$ to have a first moment $\beta I$ of order $\beta$ or larger, to persist at distances $\tau\gtrsim \beta^{-1}$ from the shock even near the grazing angle, and to induce downstream spatial variation of order $\beta^0$ in the isotropic PDF component $\phi_0$, of order $\beta^0$ in the positive and negative fluxes $j^+$ and $j^-$, and thus of order $\beta^1$ in $\myPesc(\tau)$.
It may be a-priori unclear if there are diffusion functions $D(\mu)$ that can in fact lead to such a behavior.
However, our numerical results, illustrated in Figure \ref{fig:LA}, indicate that indeed, anisotropic diffusion functions can satisfy all these criteria and yield substantial deviations from $p=p_0$, for an arbitrarily small $\beta$.

The figure demonstrates (bar-triangles, with a dot-dashed line to guide the eye) that for the simple choice $D_u=D_d=\exp(\alpha\tilde{\mu})$, the spectrum becomes exceedingly hard --- and increasingly different from $p_0$ --- as the constant $\alpha$ become negatively large.
We choose this simple exponential form of $D$ because it leads to a simple scaling of the spectrum, approaching a function of $\beta^{0.8}\alpha$ for small $\beta$.
Assuming that this behavior persists for arbitrarily small $\beta$, one can always find a sufficiently negative $\alpha$ that yields an order unity deviation from $p_0$.
Note that this specific example, while useful for demonstrative purposes, requires variations in $D(-1<\mu<1)$ that are exponential in $\beta^{-1}$, and so become non-physically large in the small $\beta$ limit.
However, one can identify other diffusion functions, with modest variations in $D$, that nevertheless produce order unity deviations in the spectrum.

One such family of angular diffusion functions, involving a factor $\sim\beta$ suppression of $D$ in a narrow beam around the forward, $\mu\simeq +1$ direction, are sufficient for a substantial hardening of the spectrum.
Curiously, the PDF shows in this case an order-unity suppression inside the beam, $\beta\,\delta\phi(\mu\simeq +1)\simeq -1$.
This demonstrates that locally, an order unity anisotropy is possible.
Moreover, if scattering is associated with modes driven by the accelerated particles themselves, it may be possible to find a self-consistent solution for the shock structure, with energetic particles missing in the forward beam responsible for their own suppressed diffusion and hard spectrum.
For a discussion of correlations between the PDF and the diffusion function, and the resulting spectral changes, see \citet{NagarKeshet19}.
We note that in spite of the strong anisotropy, the mean energy gain here is still given by Eq.~(\ref{eq:Gain}).

\section{Summary and discussion}
\label{sec:Discussion}

We have shown that for a general scattering function of particles accelerated in a non-relativistic shock, one cannot self-consistently arrive at the standard spectral index $p_0$ of Eq.~(\ref{eq:DSAp}) using previous methods; namely, by invoking the approximation of spatial diffusion on both sides of the shock, or by a priori neglecting corrections due to the PDF anisotropy.
Requiring spatial diffusion near the shock and continuity across it imposes the non-trivial consistency requirement (\ref{Eq:ScndAnisoSpatDiff}), so the spatial diffusion term must be supplemented by streaming and anisotropy-driven diffusion terms; see Eq.~(\ref{eq:transport4}).
Downstream anisotropies can be first or even zeroth order in $\beta$, and will modify the spectrum unless $\langle \mu\phi\rangle=O(\beta^2)$.

Indeed, we demonstrate in Figure \ref{fig:LA} that, in contrast to previous claims, anisotropic scattering functions can substantially modify the spectrum, even for small-angle scattering and an arbitrarily small $\beta$.
It is unclear if the spectrum can noticeably deviate from Eq.~(\ref{eq:DSAp}) in any astronomical non-relativistic shock, under the standard DSA assumptions outlined below.
Solving self-consistently for the scattering function is beyond our present capabilities, but we point out that a hard spectrum can develop if $D$ strongly correlates with $\phi$ (see \S\ref{sec:DiffExamples}).

We recover the standard spectrum (\ref{eq:DSAp}) when a suitable, $-\beta_u^{-1}\ll \tau\ll \beta_d^{-1}$  matching layer exists or when the non-diffusive terms added in Eq.~(\ref{eq:transport4}) are negligible; the result $p=p_0+O(\beta)$ is rigorously proved for an isotropic medium.
The spectrum is directly related to the shock front anisotropy, the downstream homogeneity, and the lingering of anisotropy downstream, through Eqs.~(\ref{eq:pFromAnisotropy}), (\ref{eq:pFromEvolution}), (\ref{eq:DeltaP}), and (\ref{eq:DownstreamEv}), as illustrated in Figure \ref{fig:DiffusionExample}.
The spectral index noticeably deviates from $p_0$ if downstream, the PDF anisotropy has a first moment, $\langle \mu\phi\rangle=\beta I/2$, of order $\beta$ or larger, persists at distances $\tau\gtrsim \beta^{-1}$ from the shock even near the grazing angle, and induces downstream spatial variations of order $\beta^0$ in the isotropic PDF component $\phi_0$, of order $\beta^0$ in the fluxes $j^+$ and $j^-$, and thus also of order $\beta^1$ in $\myPesc(\tau)$.

The typical assumptions underlying such DSA studies are quite strong, and include:
(i) a non-relativistic, planar shock with a well-defined, energy-independent jump in $\beta$ over a scale much shorter than the Larmor radius of the accelerated particles;
(ii) a steady state PDF, that can be averaged over constant $z$ planes to yield an effective, low dimensional PDF $f(z,q,\vect{\hat{q}})$;
and
(iii) scattering modes stationary in the fluid frame, with slowly-varying properties, that may be described using a similarly averaged scattering function $\MyScat(z,q,\vect{\hat{q}},\vect{\hat{q}}')$, with a separable angular behavior.
Note that, as the scattering function here is considered to be prescribed, the test-particle approximation is not strictly-speaking invoked.

The above assumptions could break down in many ways, with significant consequences for the spectrum.
For example, the scattering modes could be moving with respect to the fluid; a $1\lesssim\tau\lesssim\beta^{-1}$ region with different mode velocities would suffice to distort the spectrum.
An energy-independent discontinuity in the mode velocity may change the results simply by modifying the effective value of $r$.
However, a more complex shock structure, which may be generated by the backreaction of the accelerated particles, could severely modify the spectrum, as suggested by non-linear DSA studies \citep[\eg][]{drury_83, Blandford_Eichler_87, JonesEllison91, MalkovDrury01}.

\acknowledgements
We thank E. Waxman, Y. Nagar, E. Sobacchi, K.-C. Hou, and Y. Gal for helpful discussions.
This research has received funding from the GIF (grant I-1362-303.7/2016) and the IAEC-UPBC joint research foundation (grant No. 300/18), and was supported by the Israel Science Foundation (grant No. 1769/15).

\bibliography{DSA}

\end{document}